\documentclass{aa}
\usepackage{latexsym}
\usepackage{graphicx}
\usepackage{times,amssymb,verbatim}
\usepackage{natbib}
\usepackage{psfrag}
\usepackage[mathcal]{euscript}
\usepackage{mathrsfs}
\usepackage{amsmath}
\usepackage{amsbsy}
\bibpunct{(}{)}{;}{a}{}{,} 

\def\rev#1{{\bf  #1}}

\begin{document}
\title{Cygnus X-3 and the problem of the missing Wolf-Rayet X-ray binaries}

\author{Dave Lommen\inst{1,2}
	\and Lev Yungelson\inst{1,3,4}
	\and Ed van den Heuvel\inst{1}
	\and Gijs Nelemans\inst{5}
	\and Simon Portegies Zwart\inst{1,6}
	}

\institute{Astronomical Institute ``Anton Pannekoek'', University of Amsterdam,
and Center for High Energy Astrophysics, Kruislaan 403, 1098 SJ Amsterdam, The
Netherlands
\and Sterrewacht Leiden, Niels Bohrweg 2, 2333 CA Leiden, The Netherlands
\and Institute of Astronomy of the Russian Academy of Sciences, 48 Pyatniskaya
Str., 119017 Moscow, Russia
\and Isaac Newton Institute, Moscow Branch, 12 Universitetskii Pr., Moscow, Russia
\and Astronomy Department, IMAPP, Radboud Universiteit Nijmegen, Toernooiveld 1, 6525
ED Nijmegen, The Netherlands
\and Informatics Institute, University of Amsterdam, Kruislaan 403, 1098 SJ,
Amsterdam The Netherlands
}

\offprints{Dave Lommen,\\ \email{djplomme@science.uva.nl}}

\date{Received ?? / Accepted ??}

\abstract{
Cygnus X-3 is a strong X-ray source ($L_X \approx 10^{38}$ erg s$^{-1}$) which
is thought to consist of a compact object, accreting matter from a helium star.
We find analytically that the estimated ranges of mass-loss rate and orbital-period 
derivative for Cyg~X-3 are consistent with two models: i) the system is detached and the mass loss from the system comes from the stellar wind of a
{\em massive} 
helium star, of which only a fraction that allows for the observed X-ray 
luminosity is accreted, or ii) the system is semidetached and a Roche-lobe-overflowing 
{\em low- or moderate-mass} helium 
 donor transfers mass to the compact object, followed by ejection of
its excess over the Eddington rate from the system. These analytical results appear
to be consistent with evolutionary calculations. By means of population synthesis we
find that currently in the Galaxy there may exist $\sim 1$ X-ray binary with a
black hole that accretes from a $\gtrsim 7 M_\odot$ Wolf-Rayet star and $\sim 1$
X-ray binary in which a neutron star accretes matter from a 
Roche-lobe-overflowing helium star with mass $\lesssim 1.5 M_\odot$.
Cyg~X-3 is probably one of these systems.

\keywords{accretion, accretion disks -- stars: individual: (Cyg X-3) -- stars:
binaries: close -- X-rays: binaries -- X-rays: stars -- stars: Wolf-Rayet}}

\newcommand{\ud}{\mathrm{d}}
\newcommand{\uyr}{\mathrm{yr}}

\authorrunning{Dave Lommen et al.}
\titlerunning{Cyg X-3: the only WR X-ray binary}
\maketitle

\section{Introduction}
\label{sec:introduction}

Cygnus X-3 was discovered as an X-ray source by \citet{giacconi:1967}.
It is a strong X-ray source ($L_X \approx 10^{38}$ erg s$^{-1}$, assuming a distance of 9 kpc), and the X-rays are expected to be due to accretion of matter onto a compact
object (c.o.), presumably a black hole (BH) or a neutron star (NS) \citep[see, e.g.,][]{kitamoto:1987, predehl:2000}.
The X-ray and infrared (IR) emission show 
a periodicity of 4.8 hours, which is
believed to be the orbital period $P$ of the system \citep[see, e.g.,][]{parsignault:1972}. Van den Heuvel \& de Loore
(1973) suggested that Cyg X-3 consists of a NS 
with a helium (He) star companion,
as a later evolutionary product of a high-mass X-ray binary. \citet{ty73}
independently considered a NS accompanied by a He
star as a
stage in an evolutionary scenario leading from a pair of main-sequence stars to a binary
NS. There is too much interstellar obscuration towards the source to
observe it optically, but observations in the 
IR wave bands in the
1990's by van Kerkwijk and coauthors (1992, 1993, 1996) identified \nocite{vankerkwijk:1992, vankerkwijk:1993, vankerkwijk:1996}
a Wolf-Rayet (WR) spectrum with Cyg X-3.
Both the observations of van Kerkwijk and coauthors as well as high-resolution
spectroscopy by \citet{fender:1999} revealed hydrogen depletion of the mass donor.
Furthermore, phase-to-phase variations in the X-ray spectra can be explained by a
strong (factor 10-100) overabundance of carbon, nitrogen, or oxygen \citep{terasawa:1994},
consistent with a classification of the Cyg-X-3 companion as a WN-type star. This added
credibility to \citeauthor*{vandenheuvel:1973}'s and \citeauthor*{ty73}'s
prediction.\\
\\
The aims of this paper are twofold.\\
(1) We will determine what combinations of stellar masses and what mass-transfer/mass-loss mechanisms are consistent with the observed $\dot{P}/P$ values and observed mass-loss rates $\dot{M}$ from the Cyg-X-3 system (the observations are taken from the literature and summarised in Sect.~\ref{sec:Models_Introduction}). For this we make analytical estimates and carry out evolutionary calculations for systems consisting of a He star and a 
c.o. (Sects.~\ref{The models} and \ref{sec:evolutionary}).\\
(2) We will study how many X-ray sources with these (Cyg-X-3-like) 
parameters are expected to exist in the Galaxy at present.
To do this, we will carry out a population synthesis for
He-star plus c.o. (He+c.o.)
binaries in the Galaxy and 
apply disk-formation criteria to estimate the
number of observable X-ray binaries with Cyg-X-3-like parameters in
the Galaxy (Sect.~\ref{sec:population}). We
try to explain why Cyg X-3 is the only X-ray binary with a He-star companion
we observe in our galaxy 
(Sect.~\ref{sec:missing}), and briefly discuss two other binary systems that
were recently suggested to consist of a c.o. and a WR star (Sect.~\ref{sec:other}).
A Conclusion follows in Sect.~\ref{sec:Conclusions}.

\section{Observed and inferred properties of Cygnus~X-3}
\label{sec:Models_Introduction}

\subsection{Masses of the components}
\label{sec:Masses} 

\citet{tavani:1989} and later \citet{mitra:1996, mitra:1998} argued
that the  companion in Cyg X-3 should be a very-low-mass ($0.01 - 0.03 M_\odot$) He star. However, this is hard to reconcile
with the high IR luminosity of the system \citep{hanson:2000}.
Our calculations 
(Sect.~\ref{sec:Mass_transfer} \&
  \ref{sec:evolutionary}) and population synthesis
  (Sect.~\ref{sec:population}) also exclude a donor less massive than
  $\sim 0.8 M_\odot$.

\citet{terasawa:1994} found, from the ionisation structure of the
wind in Cyg~X-3, that the mass of the wind-supplying component has to
be moderate: $7^{+3}_{-2} M_\odot$.

\citet{schmutz:1996} conclude that the variations in the profiles of
several near-IR emission lines are due to the orbital motion of the WR
star and derive a mass function for the donor
$f(m_{\rm d})=2.3\,M_\odot$. For the range of assumed Wolf-Rayet masses 5 to 20
$M_\odot$ and a range of possible inclinations $30^\circ \leq i \leq
90^\circ$ they get a mass in the range $(7 - 40)   M_\odot$ for the
c.o. from which they infer that it is a BH.

\citet{hanson:2000} found a sinusoidal absorption feature originating in the wind in the 2.06$\mu$m spectral region of Cyg~X-3 which allowed them to derive a mass function 
$f(m)=0.027\,M_\odot$. Hanson et al. considered two options: an origin of the absorption in the accretion disk or other material centered on the compact object, or 
association of absorption with the donor. The first option is
consistent with low- or moderate-mass 
 ($\lesssim 8\,M_\odot$) donors, but requires a low orbital inclination of the 
system
($\lesssim 20^\circ$).
Association of the absorption feature with the donor limits the mass of the WR companion to $\lesssim 10 M_\odot$ if the accretor is a NS. For BH accretors
the mass of the
secondary may be as high as $70 M_\odot$.

\citet{stark:2003} studied the modulation of X-ray emission lines from Cyg~X-3. Based on a discussion of the location of the regions of emission of highly-ionised silicon, 
sulfur, and iron, they assume that the iron line is produced in the wind captured by the c.o. or in an accretion disk around it. They then use the fact of 
non-detection of a modulation of the iron lines to derive an upper
limit to the mass function for the accretor: $f(m_{\rm a}) \le 0.22 M_\odot$. 
For an accretor of $1.4\,M_\odot$ or $10\,M_\odot$, the minimum
  mass of the donor is then $\sim 1.1\,M_\odot$ and $\sim
  3.4\,M_\odot$, respectively. We furthermore note that in the case of Roche-lobe
  overflow (RLOF) a mass ratio $m_{\rm d}/m_{\rm a} > 1.39$ is
  inconsistent with the observed increase of the period (see
  below), which then also implies an inclination $\gtrsim 20^\circ$ in
  this case.

The early models of Cyg~X-3 that assumed an elliptic orbit for the system \citep{ghosh:1981} may be discarded now, since no signs of an apsidal motion were found in
$\simeq 30$ yr of observations \citep{singh:2002}. An implication of this non-discovery of an apsidal motion is the irrelevance of the values of the orbital inclination found by
\citet{ghosh:1981}, which are often used in the literature.

To summarise, a great ambiguity still exists in the interpretation of
the radial-velocity curves of Cyg~X-3, mostly related to different locations of
spectral features that serve as the basis for radial-velocity determinations.
However, at the moment it seems likely that \citet{stark:2003} really
measure emission originating in the vicinity of the c.o. Then their
results suggest a rather moderate mass for the companion to the c.o., if the c.o. is a neutron star or a stellar-mass black hole.

\subsection{Orbital period and its derivative}
\label{sec:Orbital_period}  

The period $P$ of Cyg X-3 has been extensively monitored over the years
\citep[e.g.,][]{vanderklis:1981, kitamoto:1987}, and is found to be increasing
on a relatively short time scale of $\sim$\hspace{1mm}$10^{6}$ yr. 
There are also indications of a second derivative of the order of
  $-10^{-10}$ yr$^{-1}$ to the period \citep{vanderklis:1989,
    kitamoto:1995}. A summary of the estimates of $\dot{P}/P$ for
  Cyg~X-3 is presented in Table~\ref{tab:Period}.
 \begin{table}
  \caption[2a]{The values for $\dot{P}/P$ of Cyg X-3, derived by fitting two different
models to the observations. A parabolic ephemeris assumes $\ddot{P} = 0$, whereas in the cubic ephemeris also a second derivative of the period unequal to zero is taken
into account.
}
  \begin{center}
   \footnotesize
   \vskip 0.2cm
   \label{tab:Period}
   \begin{tabular}{lcl}
    \hline
    \hline
    $\dot{P}/P$, fitted value                   & $\chi^2_{\rm red}$ & Reference                          \\
    \hspace{0.3cm}($10^{-6}$ yr$^{-1}$)         &                    &                                    \\
    \hline
    \multicolumn{3}{c}{Parabolic ephemeris}		                   	                                                       \\
    \hline
    $2.2 \pm 0.3$\hspace{0.1cm}$^\dagger$  	& 1.41               & \citet[][]{vanderklis:1981}        \\
    $2.19 \pm 0.05$\hspace{0.1cm}$^\dagger$	& 0.78               & \citet[][]{kitamoto:1987}          \\
    $1.6 \pm 0.1$	                        & 1.55               & \citet[][]{vanderklis:1989}        \\
    $1.2 \pm 0.4$	                        & 2.07               & \citet[][]{kitamoto:1995}          \\
    $1.05 \pm 0.04$	                        & 3.08               & \citet[][]{singh:2002}             \\
    \hline
    \multicolumn{3}{c}{Cubic ephemeris}			                   	                                                       \\
    \hline
    $4.0 \pm 0.6$\hspace{0.1cm}$^\dagger$	& 1.36               & \citet[][]{vanderklis:1989}        \\
    $2.9 \pm 0.2$\hspace{0.1cm}$^\dagger$	& 1.46               & \citet[][]{kitamoto:1995}          \\
    $1.4 \pm 0.3$\hspace{0.1cm}$^\dagger$	& 2.96               & \citet[][]{singh:2002}             \\
    \hline
   \end{tabular}
  \end{center}
$\dagger$ --  value not quoted in the referred-to paper, but calculated by us from the published values of 
$P$ and $\dot{P}$.
 \end{table}

\subsection{Mass-loss rate}
\label{sec:Mass-loss_rate} 

The mass-loss rate for Cyg X-3 was  
estimated from IR observations, usually using the
\citet[][W \& B]{wright:1975} model for the emission of a spherical,
homogeneous, constant-velocity, isothermal wind. Stars have an
accelerating wind with a temperature gradient, but W \& B note that observations show 
spectrum flattenings in the near-IR similar to those predicted by their
constant-temperature, constant-velocity model.

\citet[][]{waltman:1996} and \citet[][]{miller-jones:2005} use 
another method to estimate $\dot{M}$. 
They also assume the mass outflow to be spherically symmetric and
then use the fact that a (post-outburst) jet becomes observable with different delays after the burst at different frequencies \citep[see][~for details]{waltman:1996}. 
Note that this method gives $\dot{M}$ in the wind, {\em not} $\dot{M}$ in the jet itself, which is assumed to be much smaller.

The estimated mass-loss rates for
Cyg X-3, varying from $0.5 \times 10^{-6} M_\odot$ yr$^{-1}$ up to $2.9
\times 10^{-4} M_\odot$ yr$^{-1}$, are presented in Table~\ref{tab:Mdot}.
All estimates except one assume spherical symmetry. Note
that deviations from spherical symmetry will most probably result in a {\em higher}
mass-loss rate in the estimates from time delays \citep{waltman:1996, miller-jones:2005},
whereas deviations from spherical symmetry in the other cases will result in a
{\em lower} effective mass-loss rate from the system \citep[see, e.g.,][]{koch-miramond:2002}.

If the increase of the orbital period is considered to be
the result of a high-velocity
wind from the system that takes away specific angular momentum of the
donor 
\citep[e.g.,][]{kitamoto:1995, ergma:1998} the formula
$\dot{P}/(2 P) = - \dot{M}/M_{\rm t}$ yields $\dot{M} \approx 5 \times
10^{-7} (M_{\rm t}/M_\odot) M_\odot$ yr$^{-1}$, where $M_{\rm t}$
is the total mass in the system.

\begin{table}
  \caption[2a]{Values for $\dot{M}$ from the literature. We only show values
that are obtained from observations of the mass loss from the system, hence not
those inferred from the evolution of the orbital period. All estimates except
\citet{ogley:2001}$^\mathrm{d}$ assume spherical symmetry. See
the main text for details.
}
 \begin{center}
  \footnotesize
  \vskip 0.2cm
  \label{tab:Mdot}
  \begin{tabular}{ll}
   \hline
   \hline
   Estimated $\dot{M}$                                  & Reference             
                       \\
   \hline
   $(0.2 - 2.7) \times 10^{-5} M_\odot$ yr$^{-1}$  &
\citet{waltman:1996}$^{\mathrm{a}}$           \\
   $4 \times 10^{-5} M_\odot$ yr$^{-1}$		        &
\citet{vankerkwijk:1993}$^{\mathrm{b}}$	\\
   $\lesssim 10^{-4} M_\odot$ yr$^{-1}$        	        &
\citet{vankerkwijk:1996}$^{\mathrm{c}}$	\\
   $(0.4 - 2.9) \times 10^{-4} M_\odot$ yr$^{-1}$       & \citet{ogley:2001}	
                \\
   $\lesssim 10^{-5} M_\odot$ yr$^{-1}$		        &
\citet{ogley:2001}$^{\mathrm{d}}$		\\
   $\sim 1.2 \times 10^{-4} M_\odot$ yr$^{-1}$          &
\citet{koch-miramond:2002}$^{\mathrm{e}}$     \\
   $(0.5 - 3.6) \times 10^{-6} M_\odot$ yr$^{-1}$  &
\citet{miller-jones:2005}$^{\mathrm{f}}$      \\
   \hline
  \end{tabular}
 \end{center}
 \begin{list}{}{}
  \item[$^{\mathrm{a}}$] From delays between the 15, 8.3, and 2.25 GHz radio
light curves, assuming a jet velocity of 0.3 $c$.
  \item[$^{\mathrm{b}}$] W \& B model, from K-band observations, taking the wind
velocity to be $v_w = 1000$ km s$^{-1}$.
  \item[$^{\mathrm{c}}$] W \& B model, from new I- and K-band observations which gave
an improved value for the wind velocity $v_w \sim 1500$ km s$^{-1}$.
  \item[$^{\mathrm{d}}$] Adopting the non-spherical, disk-like model by
\citet[][]{fender:1999} and using the \citet{gorenstein:1975} approximation for
X-ray absorption.
  \item[$^{\mathrm{e}}$] W \& B model, taking $v_w = 1500$ km s$^{-1}$.
  \item[$^{\mathrm{f}}$] From delays between the 43 and 15 GHz radio light
curves, assuming a jet velocity of 0.6 $c$.
 \end{list}
\end{table}

\section{The models}
\label{The models}

We consider two possible mechanisms that may cause variations of the orbital
period in a binary system consisting of a c.o. and a companion.
In the first model
the companion loses mass in a wind which is directly lost from the system,
except a tiny fraction that may be accreted (we assume accretion at the Eddington limit). This is the model assumed by, e.g., \citet[][]{kitamoto:1995} 
and \citet[][]{ergma:1998}, see previous Section. In the second model the companion transfers mass to the c.o., which then may 
eject (part of) the transferred mass from the system. 
A similar qualitative model was first suggested for Cyg~X-3 by \citet{vanbeveren:1998}
and later for both SS~433 and Cyg~X-3 by \citet{fuchs:2002b, fuchs:2004}. In this model
the transferred matter forms an envelope around the c.o., resembling a stellar atmosphere.
A small thin accretion disk may be present around the c.o. The envelope is
ionised by X-ray emission from the vicinity of the c.o. and expelled from the system by radiation pressure.
The whole process mimics the formation of a WR-star wind and consequently of a WR-like
spectrum. Note that a disk-like wind from the system would give the same observed
spectrum as a direct spherical wind from the donor. Following \citet{fuchs:2002b, fuchs:2004}
one may speak about a ``WR phenomenon'' in this case. The line-emission region may then
really be associated with the c.o. as suggested by, e.g., \citet{fender:1999} and 
\citet{stark:2003}. The details of this qualitative picture remain to be elaborated upon and
verified by observations.

In principle the observed mass loss may be a combination of a direct wind and
re-ejection of transferred mass. 
However, since in the case of  Roche-lobe overflow (RLOF) the mass-loss rate considerably exceeds the
mass-loss rate of a direct wind (see Sect.~\ref{sec:Mass_transfer}), we
consider only the extreme cases, one being only wind mass loss (with only so
much mass transfer as to allow the c.o. to accrete at the Eddington
rate), the other being no direct wind and only mass transfer from the He
star followed by re-ejection from the compact star.

\subsection{Equation for the period derivative}
\label{sec:period}
We consider a binary system containing a mass-losing star (the donor) with mass
$m_{\rm d}$ and a c.o. (the accretor) with mass $m_{\rm a}$. The
donor loses an amount of mass $\ud m_{\rm d}$ of which a
fraction $\alpha$ is directly lost from the system, carrying away the specific
orbital angular momentum of the donor. A fraction $\beta$ is first transferred
to the accretor and then lost through re-ejection, carrying away $\beta \ud m_{\rm d}$
in mass with the specific orbital angular momentum of the accretor.
We can then find $\dot{P}/P$ as a direct function of $\alpha$, $\beta$, $m_{\rm d}$,
$m_{\rm a}$, and $\dot{m}_{\rm d}$ (see Appendix~A):
  \begin{align}\label{eq:dp_P}
    & \frac{\dot{P}}{P} = \frac{\dot{m_{\rm d}}}{m_{\rm a} m_{\rm d} (m_{\rm a} + m_{\rm d})} \times \nonumber\\
    \times & \left[ (3 m_{\rm a}^2 - 2 m_{\rm a} m_{\rm d} - 3 m_{\rm d}^2)\alpha - 2 m_{\rm a} m_{\rm d} \beta + 3 m_{\rm d}^2 - 3 m_{\rm a}^2 \right].
  \end{align}
We assume the c.o. to accrete at the Eddington rate, which leaves four
free parameters to deal with to obtain a time scale for orbital-period
variations in the range of observed values for $\dot{P}/{P}$ (see
Table~\ref{tab:Period}). 
Note that the cubic ephemeris ($\ddot{P} \neq 0$, in all those cases $\ddot{P} <
0$) fit the observations (slightly) better than the parabolic ephemeris.
Although the cubic ephemeris of \citet[][]{vanderklis:1989} and
\citet[][]{kitamoto:1995} are not confirmed --- according to both ephemeris
the period $P$ should have stopped increasing and started {\em decreasing} by
now, which is not observed --- 
we note that $\ddot{P} < 0$ may
indicate a decrease in the mass-loss rate. Discarding the not-confirmed
cubic models c and d in Table~\ref{tab:Period}, the observed $\dot{P}/P$ values
are in the range $1.0 \times 10^{-6}$ yr$^{-1}$ to $2.2 \times 10^{-6}$
yr$^{-1}$. We note here that in the extreme case in which there is only re-ejection
and no direct WR wind (i.e., $\alpha = 0, \beta = 1$), only ratios 
of $m_{\rm d}/m_{\rm a} < 1.39$ give an increase of the orbital period
and are thus consistent with the observed $\dot{P}$ of Cyg X-3.

\subsection{Limits on the component masses derived from the rate of mass
transfer and mass loss}\label{sec:Mass_transfer}

\subsubsection{Mass loss due to a direct wind}\label{sec:Wind_accretion}
\begin{figure}
	\psfrag{Helium-star mass (MSun)}[t][c][3][0]{{\sf M$_{\sf He}$/M$_\odot$}}
	\psfrag{Mass-loss rates of helium stars}[t][c][3][0]{}
	\psfrag{log mass-loss rate (MSun/yr)}[t][c][3][0]{{\sf log(dM/dt) M$_\odot$/yr}}
	\resizebox{\hsize}{!}{\includegraphics[angle=270]{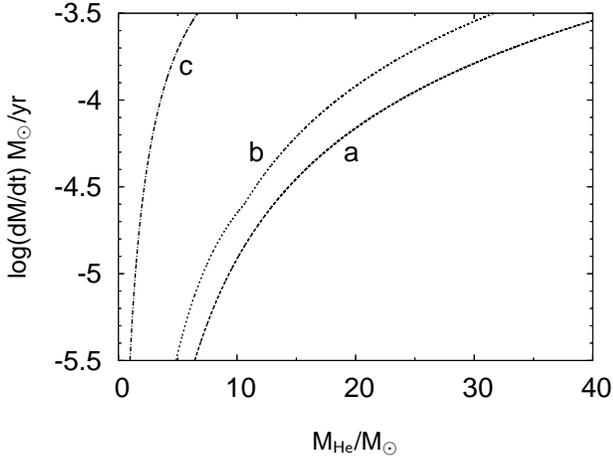}}
	\caption{ Mass-loss rates of He stars: (a) -- wind mass-loss rate for a
homogeneous He star;  (b) -- wind mass-loss rate for a He star at He terminal-age
main sequence;  (c) -- rough estimate of the mass-loss rate of a star that overflows
its Roche lobe on a thermal time scale after completion of core He
burning.} 
	\label{fig:mass-loss}
\end{figure}

The presence of relativistic jets \citep[e.g.,][]{mioduszewski:2001} shows that
there must be an accretion disk around Cyg X-3. This accretion disk may be
either formed through wind accretion or through RLOF. We will
first consider the case where the He star loses mass in a wind and the compact
object accretes part of this wind (presumably at the Eddington rate). 
For the wind mass-loss rate of a He star we use 
\begin{equation}
 \label{eq:massloss-helium}
        \dot{M} = 
	\left\{
	\begin{array}{ll}
	  2.8 \times 10^{-13} (L/L_\odot)^{1.5} {\bf \textrm{~~if} }  & \log{L/L_\odot} \ge 4.5 \\
	  4.0 \times 10^{-37} (L/L_\odot)^{6.8} {\bf \textrm{~~if} }  & \log{L/L_\odot} < 4.5
	\end{array}
	\right.
\end{equation}
\citep[in $M_\odot$ yr$^{-1}$, see][~and references therein]{dewi:2002}. 
We take
the luminosity of a He star on the He main-sequence (HeMS) from
\citet{hurley:2000}. 
\citet{hurley:2000} only did calculations for He stars up to $10 M_\odot$, but \citet[][]{nugis:2000} find that when
\citeauthor{hurley:2000}'s results are extrapolated up to He-star masses of $M_{\rm He} = 40 M_\odot$ they are consistent with the results of
\citet{schaerer:1992}, which are valid for WNE\footnote{early WN stars, WN2-5}/WC stars up to $40 M_\odot$.
In Fig.~\ref{fig:mass-loss} we plot the thus obtained mass-loss rate for a He
star as a function of its mass $M_{\rm He}$
at the beginning and at the end of the HeMS.

Observed population I WR stars are more massive than $\sim 7 M_\odot$ \citep[see,
e.g.,][]{nugis:2000}; lower-mass He stars will probably not show a WR spectrum,
nor will they produce a wind that could explain the mass-loss rate observed in
the Cyg-X-3 system.
\citeauthor[][]{nugis:2000} do not report any WR stars with mass-loss rates below $4 \times 10^{-6} M_\odot$ yr$^{-1}$, whereas \citet[][]{miller-jones:2005} find that the 
mass-loss rate from Cyg X-3 {\em may} be below $4 \times 10^{-6} M_\odot$ yr$^{-1}$. This might be an indication that the WR spectrum observed from Cyg X-3 is due to 
the re-ejection of Roche-lobe-overflowed material that mimics the WR phenomenon. Another explanation, however, may be that the outflow
is not spherically symmetric in the case of Cyg X-3 (Sect.~\ref{sec:Models_Introduction}).

\subsubsection{Mass transfer due to RLOF}\label{sec:RLOF_accretion}

Evolutionary calculations of \citet[][]{paczynski:1971}, \citet[][]{ty73.1}, and \citet[][]{iben:1985}
showed that He stars with $M_{\rm He} \lesssim 0.8 M_\odot$ do not
expand during core He burning and later evolutionary stages.
Also He stars with $M_{\rm He} \gtrsim 8 M_\odot$ hardly expand before carbon (C) ignition in their cores (later stages are so short that they may be neglected).
For intermediate-mass He stars, inspection of the summary figure of
\citet[][~their Fig.~1]{dewi:2002} shows that in a binary with $P = 0.2$ days containing a NS and a 
He star,
the latter may overflow its Roche lobe in the He-shell-burning stage if $M_{\rm He} \lesssim 5.8 M_\odot$ or in the core-C-burning stage if $M_{\rm He} \lesssim 7.4 M_\odot$.
However, the expected number of systems in the Galaxy that experience RLOF in the C-burning stage is negligibly small since this stage is very 
short,
and we are left with (0.8 - 5.8)$M_\odot$ He stars, overflowing their
Roche lobe in the He-shell-burning stage (so-called BB case of evolution).
If RLOF occurs after core He burning is completed, the mass-exchange
time scale is of the order of the thermal one:
\begin{equation} \label{eq:Kelvin-Helmholtz}
	\tau_{\rm th} = \frac{G M^2}{R L},
\end{equation}
with $M$, $R$, and $L$ the mass, radius and luminosity of the
donor. 
In Fig.~\ref{fig:mass-loss} we also plot the mass-transfer rate from the He star
given by Eq.~(\ref{eq:Kelvin-Helmholtz}),
with
$R$ and $L$ at the end of
the He main-sequence from \citet[][]{hurley:2000}. 
We see
that if in a system containing a rather low-mass He star matter is transferred
to the c.o. via RLOF 
in a thermal time scale, the 
rate of loss of the re-ejected matter from the system
would cover
the same range in $\dot{M}$ as that for the winds of WR stars. Notice that
Fig.~\ref{fig:mass-loss} justifies the consideration of only the two extreme
cases of mass loss (stellar wind vs. RLOF+re-ejection): in a system with a given
He-star mass that experiences RLOF, the mass transfer rate due to RLOF is two
orders of magnitude larger than the wind mass-loss rate of a He star with the
same mass would be.

\subsubsection{Analytical results for the Cyg-X-3
system}\label{sec:analytical_results_CygX3}

\begin{figure}
	\psfrag{Donor mass (solar masses)}[t][c][3][0]{{\sf Donor mass (M$_\odot$)}}
	\psfrag{Accretor mass (solar masses)}[b][c][3][0]{{\sf Accretor mass (M$_\odot$)}}
	\resizebox{\hsize}{!}{\includegraphics[angle=270]{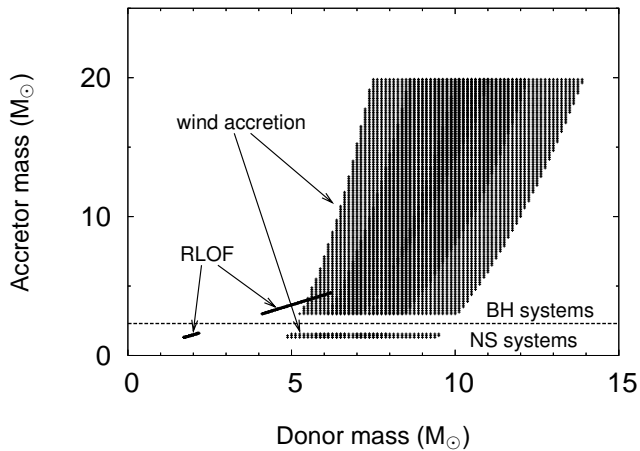}}
	\caption{Schematic representation of possible system configurations that
are consistent with $\dot{P}/P$ and $\dot{M}$ as observed in Cyg~X-3. 
RLOF solutions give a very narrow range of possible combinations of $m_{\rm a}$ and
$m_{\rm d}$ (the two 
narrow strips in the lower left; the gap is due to the presumed gap
between the masses of NSs and those of BHs). For wind-accretion solutions the
range of possible combinations of $m_{\rm a}$ and $m_{\rm d}$ is much larger (hatched areas).
Population synthesis shows that RLOF systems with NSs are formed in sufficiently
large numbers to produce an observable Cyg-X-3 system. On the other hand, Roche-lobe overflowing BH
systems are formed so rarely with $P$ similar to the period of Cyg~X-3 that the probability
of observing them is negligible. See Figs.~\ref{fig:ns_wr} and \ref{fig:bh_wr}
and text.}
	\label{fig:ma_vs_md}
\end{figure}

We can now insert the estimated values of RLOF and wind mass-loss rates given
by Eqs.~(\ref{eq:massloss-helium}) and (\ref{eq:Kelvin-Helmholtz}) and the observed
$\dot{P}/P$ into Eq.~(\ref{eq:dp_P}) and solve it for all possible combinations of
masses of the components $m_{\rm a}$ and $m_{\rm d}$ that satisfy the observations for Cyg
X-3. The result of this is shown in Fig.~\ref{fig:ma_vs_md}.
The 
narrow strips in the lower left show the possible combinations, assuming that
mass transfer in the system is due to RLOF in a thermal time scale. The strips are
shown here to exemplify their narrowness: the width of the strip is typically less than $0.05 M_\odot$.
When one uses mass-transfer rates for
RLOF obtained from real binary-stellar-evolution calculations, the position of
the strip may be somewhat displaced, but it remains narrow. The larger areas to
the right show the possible mass combinations, assuming that mass-transfer in the
system is due to wind accretion. We note that the much larger areas for the
wind-accretion solutions do not mean that wind accretion is the most probable
means of mass transfer in the Cyg-X-3 system. It merely shows that, {\em if}
wind accretion is the mass-transfer mechanism in Cyg X-3, the space of possible~
$m_{\rm a}$-$m_{\rm d}$ combinations is still quite large, whereas RLOF as the mass-transfer
mechanism leaves only few possible combinations of $m_{\rm a}$ and $m_{\rm d}$. However,
results of population synthesis (see below) show that under the assumptions
leading to the formation of He-star+BH (He+BH) systems, only a minor part of the
``allowed'' area is populated. Fig.~\ref{fig:ma_vs_md} shows that the observed
mass-loss rate from the system and the observed $\dot{P}/P$ are hard to
reconcile with the suggestions of donor masses of several $10 M_\odot$.

\section{Evolutionary calculations}\label{sec:evolutionary}
Analytical estimates for Roche-lobe-overflowing systems presented above suggest
that the ``observed'' $\dot{M}$ range of Cyg X-3 may be typical for
moderate-mass (up to
several $M_\odot$) He stars overflowing their Roche lobes after
completion of core He burning. 
The results of our population synthesis also suggest that most He companions
to compact objects are of moderate mass, see Sect.~\ref{sec:population}.

\begin{figure*}[!htp]
	\resizebox{\hsize}{!}{\includegraphics[angle=270]{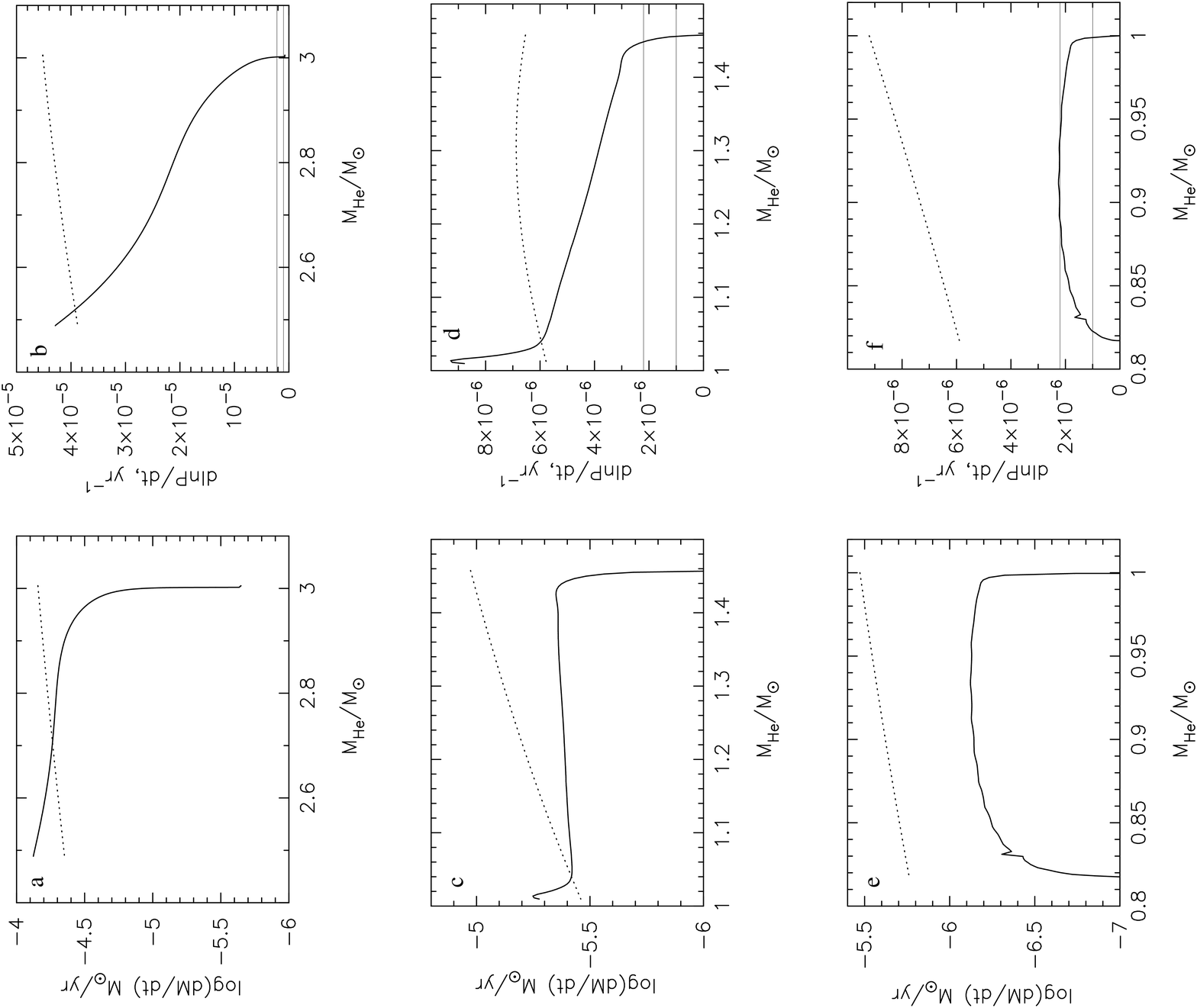}}
	\caption{ $\dot{M}$ and $\dot{P}/P$ as function of $M_{\rm He}$ for systems with
Roche-lobe overflowing He-star donors and compact accretors. 
Masses $(m_{\rm d}, m_{\rm a})$ at the onset of mass transfer are: (3.0, 5.0), (1.46, 1.4), (1.0, 1.4) $M_\odot$ (from top to bottom).
Roche-lobe overflow starts at $P \approx 0.2$ day.
In all panels thick solid lines show results of
computations.
In panels for $\dot{P}/P$ thin solid lines show
the limits of observed $\dot{P}/P$ in Cyg X-3. 
Dotted curves show rough estimates for $\dot{M}$ and $\dot{P}/P$ based on Eq.~(\ref{eq:Kelvin-Helmholtz}), derived with the approximations to $R$ and $L$ at the terminal-age
He main-sequence from \citet{hurley:2000}.
}
	\label{fig:tracks}
\end{figure*}

To verify the inferences in the previous Section
we carried out several evolutionary calculations of 
semidetached systems  consisting of
He stars accompanied by a c.o. We assumed that the
latter can accrete matter at $\dot{M} \le \dot{M}_{\rm Edd}$ and that the excess of
the transferred mass is lost from the system, taking away the specific angular
momentum of the accretor.
Prior to RLOF, mass loss by stellar wind was computed according to formulae  (\ref{eq:massloss-helium}). Accretion prior to RLOF was neglected.
In the RLOF stage wind
mass loss directly from the He star was
neglected (see Fig.~\ref{fig:mass-loss}).
The adopted ranges of initial masses were $1.0 M_\odot$--$4.1 M_\odot$
  for the He stars and $1.4 M_\odot$--$5.0 M_\odot$ for the compact objects.
Computations were carried out using P.~Eggleton's evolutionary code \citep[priv. comm. 2003, see also][~and references therein]{eggleton:2002}.
A selection of the results 
are presented graphically in Fig.~\ref{fig:tracks}.
The systems had the following combinations of component masses at
  the onset of mass transfer:
$m_{\rm d} =  3.0 M_\odot$, $m_{\rm a} =  5.0
M_\odot$ (Fig.~\ref{fig:tracks} a, b), $m_{\rm d} =  1.46 M_\odot$, $m_{\rm a} = 1.4 M_\odot$
(Fig.~\ref{fig:tracks} c, d), and $m_{\rm d} = 1.0 M_\odot$, $m_{\rm a} = 1.4 M_\odot$
(Fig.~\ref{fig:tracks} e, f).
In all the computed systems the He stars started RLOF at $P \approx 0.2$\,day.

We find that the systems with He-star donors $\gtrsim 3.0 M_\odot$ traverse the range
of $\dot{P}/P$ observed for Cyg X-3
in $\sim\,10^2$ yr. We do note, however, that the typical value of $\dot{P}/P$ in
the RLOF phase decreases with the mass of the donor (for a given mass of the compact star)
and that the time spent close to the observed $\dot{P}/P$ range
increases for lower-mass systems.
A system that at the onset of mass transfer consists of a $1.46 M_\odot$ He star and a $1.4
M_\odot$ c.o. (Fig.~\ref{fig:tracks} c, d) spends about
$5 \times 10^3$ yrs in the $\dot{P}/P$ range observed in Cyg X-3, and stays some $6 \times 10^4$ yrs at $\dot{P}/P$ values less then 
twice those in the observed range. A system 
 of a $1.0 M_\odot$ He star and a $1.4 M_\odot$ c.o. stays within the
observed range all throughout RLOF (Fig.~\ref{fig:tracks} e, f).

\section{Population synthesis}\label{sec:population}

We carried out a population synthesis to determine the current number of He+c.o. binaries in the Galaxy.
The details of the population synthesis are briefly described in Appendix B. We used the approximations of \citet{pols:1993} to the computations of \citet{paczynski:1971}
and \citet{habets:1986} to estimate the core-He-burning times. In Figs.~\ref{fig:bh_wr} and \ref{fig:ns_wr} we plot the masses of the components and the orbital periods of 
the He+c.o. systems that have He components in the core-He-burning stage. We find that there are currently $\sim$ 200 He+BH and $\sim$ 540 He+NS binaries in the 
Galaxy.\footnote{Note that these numbers represent one possible random realisation of the model for a population of He+c.o. systems, so all numbers given
are subject to Poisson noise.}

\subsection{Wind-fed X-ray systems}

As noted in Sect.~\ref{sec:Wind_accretion} we expect an accretion disk in the Cyg-X-3 system.
The systems shown in Figs.~\ref{fig:bh_wr} and \ref{fig:ns_wr} may form a disk through wind accretion if the wind matter carries enough angular momentum
as realised first by \citet[][]{illarionov:1975}; see, e.g., \citet[][]{livio:1994} for later work on this subject.
\subsubsection{Helium-star/black-hole binaries}
Following the derivation in \citet{ergma:1998} for
systems with Kerr BHs, we apply the disk-formation criterion
\begin{equation}
  P \lesssim 0.2 (M_{\rm BH}/M_\odot) v^{-4}_{1000}~{\rm day}, 
\label{eq:pcrbh}
\end{equation}
where $M_{\rm BH}$ is the BH mass, and 
$v_{1000}$ is the magnitude of the radial
velocity of the wind in the vicinity of the c.o. in units of 1000 km s$^{-1}$.
It appears then that for $v_{1000}=1$ there are currently $\sim$ 30
wind-fed disk-forming systems in the Galaxy, 5 of them with orbital periods 
similar to that of Cyg X-3.
However, only 9 out of the 30 systems have $M_{\rm He} \gtrsim 7
M_\odot$ donors and would be identified as WR stars.
Only one of these systems has an orbital period close to that of Cyg X-3.
The remaining four systems with $P$ in the observed range 
have
 $M_{\rm He} \approx (2 - 4) M_\odot$; according to
Eq.~(\ref{eq:massloss-helium}) their wind mass-loss rate
will be below $\sim 10^{-7} M_\odot$ yr$^{-1}$  and they 
probably will not produce the WR phenomenon
\footnote{A possible reason for the absence of the WR phenomenon in 
low-mass He stars is that, for luminosity-to-mass ratios characteristic for them,
one does not expect radial pulsations that may drive shock waves with a consequential increase of the gas density,
necessary to produce radiative stellar winds \citep[see, e.g.,][]{fadeyev:2003}.}.
There are 8 systems with $M_{\rm He} \gtrsim 7 M_\odot$ and $P \gtrsim 10$ hr
that fulfil the disk-formation criterion. Note, however, that
criterion~(\ref{eq:pcrbh}) has a very steep dependence on $v_{1000}$; if we assume, e.g.,
that $v_{1000}=1.5$, only 3 systems out of the 30 able to form disks remain, all
of them with $P <$ 0.15 day and with $M_{\rm He} \approx 3\,M_\odot$. Since known WR
stars have $v_{1000} \approx (0.7 - 5.0)$ \citep{nugis:2000}, we claim that Cyg X-3 may well be the only wind-fed WR+BH system in the Galaxy.

\begin{figure}
	\resizebox{0.38\vsize}{!}{\includegraphics[angle=-90,scale=0.7]{mwr_p.bh2.ps}} 
	\resizebox{0.38\vsize}{!}{\includegraphics[scale=0.7]{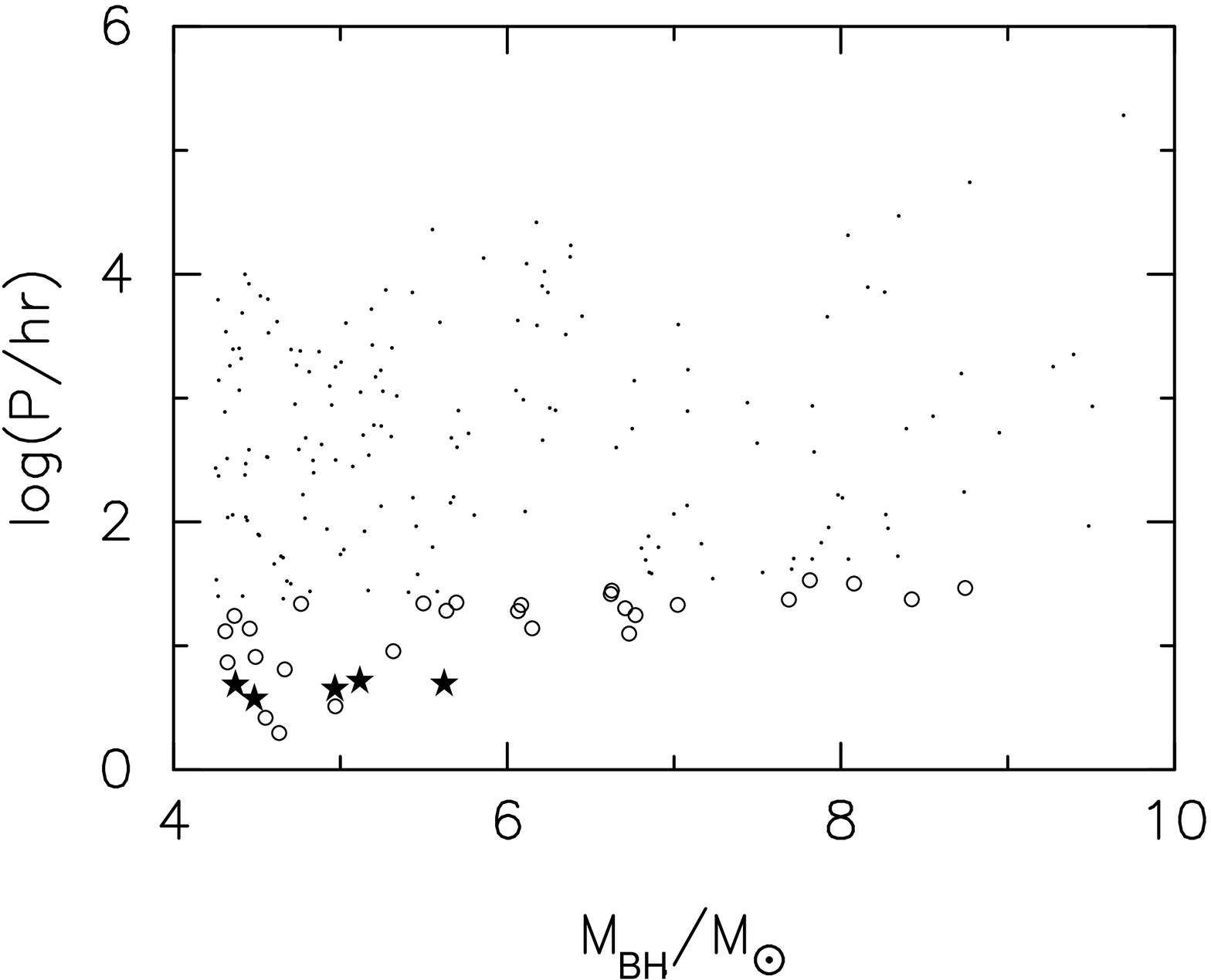}}
	\resizebox{0.38\vsize}{!}{\includegraphics[angle=-270,scale=0.7]{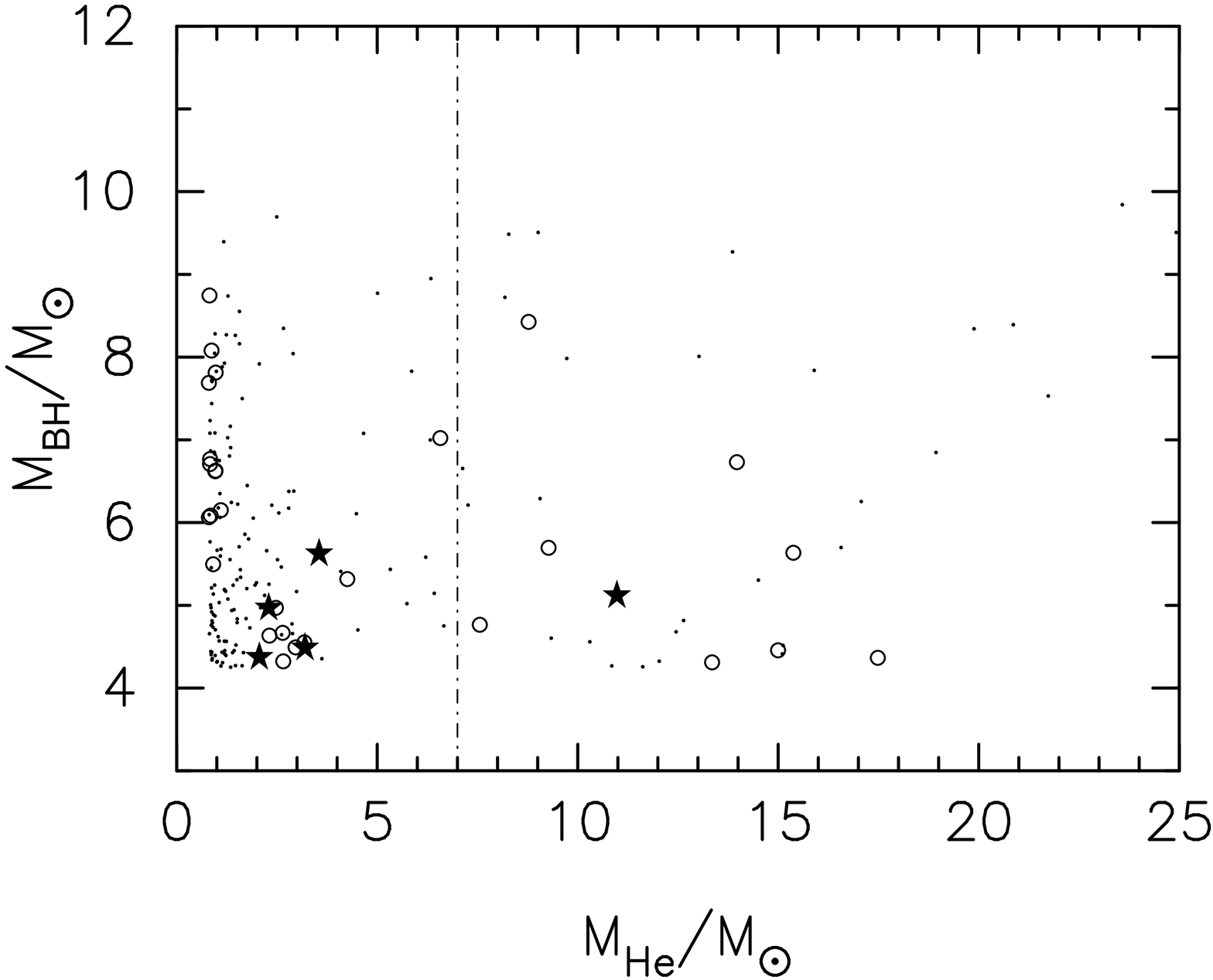}}
    	\caption{Current population of core-He-burning He+BH binaries
in the Galaxy. 
Upper panel --- distribution in $M_{\rm He} - P$ plane; middle panel ---
distribution in {\bf $M_{\rm BH} - P$} plane; lower panel --- distribution in 
{\bf $M_{\rm He} - M_{\rm BH}$}  plane. The dash-dotted vertical lines in the upper and lower
panel show the lower-mass boundary for He stars identified as WR stars.
Systems that satisfy the disk-formation criterion for wind-fed objects
(Eq.~\ref{eq:pcrbh}) are marked by open circles; the subset of them with $3.6
\leq P/{\rm hr} \leq 6.0$ is marked by stars. }
	\label{fig:bh_wr}
\end{figure}

\begin{figure}
	\resizebox{\hsize}{!}{\includegraphics[angle=270]{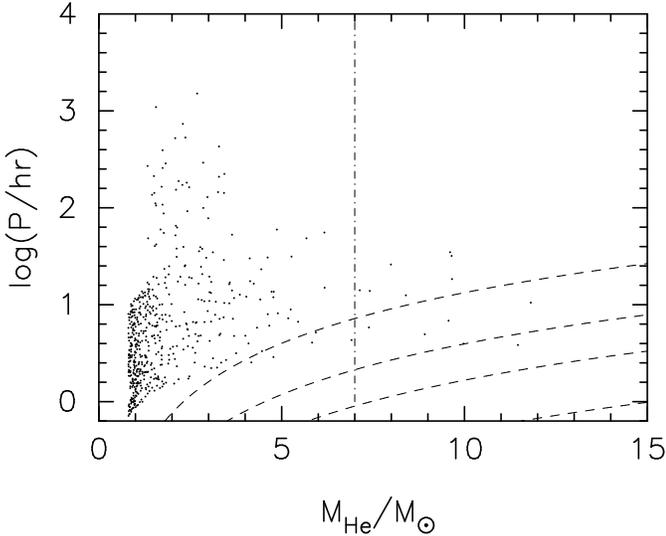}}
	\caption{Current population of core-He-burning He+NS binaries
in the Galaxy. 
The dash-dotted vertical line shows the lower-mass boundary for He stars
identified with WR stars. The four dashed lines show the critical periods below
which according to criterion (\ref{crit:ns}) disk formation through wind accretion is possible,
if $v_{1000}$ = 1, 1.5, 2, and 3 (highest to lowest), respectively.}
	\label{fig:ns_wr}
\end{figure}

As already noted by \citet{iben:1995} and \citet{ergma:1998}, with $v_{1000} \approx 1.5$,
Cyg X-3 may have a wind-fed disk if $M_{\rm BH} \approx
5 M_\odot$. The latter value fits well into the model range of expected black-hole
masses in wind-fed systems with disks and orbital periods close to that of
Cyg~X-3 (the star symbols in Fig.~\ref{fig:bh_wr}).
An additional reason why only a WR system with $P \lesssim 0.25$
days shows up as a WR X-ray binary may have to do with the velocity profile of
the wind. In such a close system the wind will, at the orbit of the compact
object, not yet have reached its terminal velocity (of for example 1500 km
s$^{-1}$), whereas in the wider systems it may have, such that no disk forms in
the wider systems. \\

\subsubsection{Helium-star/neutron-star binaries}
Figure \ref{fig:ns_wr} shows the population of 
He+NS binaries, which can be divided into two subpopulations. The first and larger subpopulation consists of systems with $P \lesssim 1$ day and $M_{\rm He} \lesssim 2 M_\odot$. The short period is due to their previous common-envelope (CE) phase; the large number of low-mass He-star systems is due to the initial-mass function and to the fact that low-mass He stars live much longer than higher-mass He stars. The other systems form through a ``double spiral-in'' in which two giants go through a CE phase, producing two He stars. This formation channel only occurs for nearly-equal-mass binaries, and since the most massive He star collapses into a c.o. there is a lower limit to the mass of the other He star \citep[see][]{brown:1995}.

As shown by \citet{illarionov:1975} and \citet{ergma:1998} an accretion disk will form in a He+NS binary if the rotation
period of the NS is longer than the equilibrium period that was
established during the CE episode that accompanied the formation of
the He star. Assuming that the wind-mass-loss rate may be described by a formula $\dot M = k M_{\rm He}^{\alpha}$ \citep[e.g.][]{nelemans:2001} one derives as the disk-formation
criterion
\begin{equation}
  P \lesssim 1.5 \cdot 10^5 k^{0.75} M_{\rm NS}^{1.5} M_{\rm t}^{-0.5} M_{\rm He}^{0.75 \alpha} v_{1000}^{-3}~{\rm hr}.
\label{crit:ns} 
\end{equation}

\noindent
[This is equivalent with Eq.~(9) of \citet{ergma:1998}, but with a
more general expression for $\dot{M}$]. We assume $k=1.38 \times 10^{-8}$,
$\alpha=2.87$ after \citet{nelemans:2001}, which gives practically the same
$\dot{M}$ values as Eq.~(\ref{eq:massloss-helium}).
Limiting periods for $v_{1000}$ = 1, 1.5, 2, and 3 are plotted in
Fig.~\ref{fig:ns_wr} (dashes). 
We 
find $\sim$ 5 systems that may be identified as WR+NS systems with accretion disks.
However, the
very steep dependence on the wind velocity may reduce this number to 0.
Thus, as already noted by \citet[][]{ergma:1998}, the low angular momentum
of WR-star winds may completely preclude the formation of wind-fed Cyg-X-3-like systems with NS companions.

One should note, however, that
the mass-loss rates for He stars quoted above
were derived from observational data on WR stars. 
Data on mass-loss rates for lower-mass He stars are not available.
It is therefore not clear
whether $\dot M$ may be extrapolated to below $\sim 7 M_\odot$. 
Hence, the validity of criterion (\ref{crit:ns}) below this mass is
uncertain.

\subsection{X-ray systems powered by RLOF}

Another possibility for Cyg X-3 is that the system contains a He star of
$\lesssim 7 M_\odot$ which transfers mass in the case BB of RLOF.
Then the 
WR spectrum arises in
re-ejected matter.
Also in the RLOF case we can estimate the number of systems we should currently
observe as WR X-ray binaries. As mentioned in Sect.~\ref{sec:evolutionary} the
RLOF systems that can provide a $\dot{P}/P$ close to the observed
range must initially have had a low-mass ($\lesssim$ 1.5 $M_\odot$) He-star donor.
From the population-synthesis calculations we expect $\sim$ 500 such systems in the Galaxy at any time.
Typically these systems live $\sim 1.5 \times 10^7$ yrs, of which they spend 
$\sim 10^5$ yrs in the phase of Roche-lobe overflow.
We thus expect only of order 1 such system in the Galaxy in the phase of Roche-lobe overflow at any time. 

As indicated by Fig.~\ref{fig:ns_wr}, the bulk of these systems has $P \le 10$
hr. The typical mass-transfer rates in
these systems are in the range $(1 - 3) \times 10^{-6} M_\odot$ yr$^{-1}$, and most of the transferred mass will be lost from the system through re-ejection.
These rates are consistent with the lowest observational estimates of $\dot{M}$ for Cyg X-3.

\section{Discussion}

\subsection{The ``missing'' He+c.o. binaries}\label{sec:missing}

Our 
population synthesis shows that there are several core-He-burning He+NS binaries and a few dozen core-He-burning He+BH binaries with He-star masses 
$\gtrsim 7 M_\odot$ in our Galaxy.
If we assume that all matter that passes through the so-called
accretion radius $r_{\rm a} = 2 G m_{\rm a} / c^2$ ($G$ the
  gravitational constant, $c$ the 
speed of light) is accreted by the c.o., and that the gravitational potential energy of the accreted matter is converted into luminosity, all wind-accreting He-star binaries 
with $M_{\rm He} > 3 M_\odot$ in our model population 
 will have an intrinsic luminosity $\gtrsim 10^{36}$ erg s$^{-1}$ and should be observable as He-star X-ray sources. 
In this we also assume that Eq.~(\ref{eq:massloss-helium}) holds down to low-mass He stars.

Apart from Cyg~X-3 a few WR+c.o. candidates are reported, e.g.,  HD~197406/WR~148 \citep{marchenko:1996},
HD~191765/WR~134 \citep{morel:1999}, HD~104994/WR~46 \citep{marchenko:2000}. However, it is still unclear whether the companions to these WR stars really have a relativistic
nature; the systems lack the X-ray luminosity expected in such case. A
low $L_X$ can be reconsiled with, e.g., a spinning pulsar, that
deflects the flow.

The fact that we do not observe several tens of He-star X-ray binaries in the Galaxy may be due to self-absorption of the X-ray photons by the wind of the donor.
It turns out that for the binaries in our population-synthesis sample the minimum column density between the c.o. and Earth due to the He-star wind 
depends mainly on $M_{\rm He}$.
This column density is $\gtrsim 10$ g cm$^{-2}$ for all sources with $M_{\rm He} > 4 M_\odot$, rendering these sources unobservable in X-rays at energies below 20 keV. We do 
note that INTEGRAL has discovered several sources at energies $> 20$ keV, of which $\sim$ 40 are still unidentified \citep[][]{ubertini:2005}. These hard sources might well 
be the missing He+c.o. binaries.

The derivation for the column density also applies to RLOF-accreting sources that spherically symmetrically throw out overflowing matter in excess to the Eddington rate. We 
saw in Sect.~\ref{The models} that the mass-transfer rate for RLOF systems is $\gtrsim 5 \times 10^{-6} M_\odot$ yr$^{-1}$, well above the Eddington rate for a solar-mass 
c.o. 
($\dot{M}_{\rm Edd} \approx 8 
\times 10^{-8} M_\odot$ yr$^{-1}$ for a 1.4 $M_\odot$ NS accreting pure He)
and good enough for a minimum column density of 
$\sim 10^2$ g cm$^{-2}$. This may support the suggestion of a model for Cyg X-3 in which the excess matter is thrown out of the system equatorially instead of spherically 
symmetrically (Sect.~\ref{The models}) together with a very low inclination for the system.

\subsection{IC10 X-1 and SS 433, two other Wolf-Rayet X-ray
binaries?}\label{sec:other}

\subsubsection{IC10 X-1}\label{sec:IC10X1}


\citet{bauer:2004} and \citet{clark:2004} find that  
the luminous X-ray source IC10~X-1 [$L_{0.1-2.5 {\rm keV}}=(2 - 4)\times 10^{38}$ erg
s$^{-1}$] 
in the starburst galaxy IC10 is spatially coincident
with  WNE star [MAC92] 17-A \citep[notation adopted from][]{crowther:2003}. 
Assuming  [MAC92] 17-A to be the most probable optical counterpart of
IC10~X-1,
\citet[][]{clark:2004} fit a model with a stellar temperature of 85\,000 K, log($L/L_\odot$) =
6.05, $\dot{M} = 4 \times 10^{-6} M_\odot$ yr$^{-1}$, and a terminal wind velocity of 1750 km
s$^{-1}$ to the observed He II $\lambda4686$ and N V $\lambda\lambda$4603--20
emission. They infer a mass for the WR star of
$\sim 35 M_\odot$, using the WR mass -- luminosity
relations of \citet[][]{schaerer:1992}.
 Allowing for clumping, \citeauthor{clark:2004}
find that $\dot{M}$ is equivalent to a homogeneous mass-loss rate of
 $\sim 1 \times 10^{-5} M_\odot$ yr$^{-1}$.

\citet{bauer:2004} note that  IC10~X-1 is quite similar to Cyg
X-3 in terms of X-ray luminosity, spectrum, and variability. Thus, IC10~X-1 may
be the first extragalactic example of a short-living WR X-ray
binary similar to Cyg X-3. The identification of the optical counterpart to IC10~X1 with a massive WR
star suggests that the system is wind-fed.\footnote{Note that the
  estimate of $\dot{M}$ quoted above is more than an order of
  magnitude lower than would be expected for such a massive WR star,
  even when keeping in mind the low metallicity of IC10: $Z/Z_\odot \approx
    0.25$ \citep[e.g.,][]{lequeux:1979}.}

Note, however, that in the field of IC10~X-1 there  are three other 
 candidate optical counterparts, with O- or B-spectral types, to the X-ray source. 
\citet{clark:2004} suggest  that their wind mass loss is
insufficient to 
explain the observed X-ray luminosity with wind accretion.
From this \citeauthor{clark:2004} argue that, if one of those three candidates
is the optical counterpart to IC10~X-1, the system is Roche-lobe-overflow fed
similar to LMC X-4 or LMC X-3. 

\subsubsection{SS 433}\label{sec:SS433}

Also the variable X-ray and radio source SS~433 was suggested to be a
WR X-ray binary \citep{heuveletal80,fuchs:2002b,fuchs:2004}. 
Van den Heuvel et al. suggested that SS~433 
contains an evolved early-type star or a WR star, based on the nature of 
its stationary spectrum, the size of the emitting region, the necessary
presence of a strong
wind for the production of the IR emission, and the large outflow velocity of the wind.

In one of the latest attempts to identify the optical counterpart of SS~433,
\citet{fuchs:2002b,fuchs:2002} compared its mid-IR spectrum to WR stars of the WN
subtype. They found the spectrum of  SS~433 to resemble
that of WR~147, a
WN8+B0.5V binary with colliding wind.
Using the formulae of \citet[][]{wright:1975}
and taking wind clumping into account, \citet{fuchs:2004} obtain 
$\dot{M}=(5 - 7)\times 10^{-5} M_\odot$ yr$^{-1}$,
compatible to a strong late-WN wind.

\citet[][]{fuchs:2004} propose that the material surrounding the c.o. forms a thick torus or envelope around it rather than a classical thin accretion disk.
They argue that the material is ionised by X-rays emitted from the vicinity of the c.o. and expelled by radiation pressure which results in the imitation of a WR star.
As mentioned above, this model needs further elaboration, especially the formation of the WR spectrum and the self-absorption of the X-rays.

On the other hand,
\citet[][]{king:2000} suggested that SS~433 is a mass-transferring system in which the formation of a CE may be avoided
if radiation pressure expels the transferred matter in excess of the
Eddington rate, i.e., a re-ejection model with a hydrogen-rich donor.
For this model the donor mass must be in the range of $(4-12) M_\odot$.
The  model received support by the discovery of
A-type super-giant features in the spectrum 
of SS~433 observable at certain orbital phases 
\citep{gies:2002,hillwig:2004}. 
Estimated masses of the components are $10.9 \pm 3.1$ and $2.9 \pm 0.7 M_\odot$,
fitting the \citeauthor{king:2000} model.
As noted by \citet[][]{fuchs:2004}, if the results of \citet{gies:2002} and \citet[][]{hillwig:2004}
are confirmed, one needs to resolve an apparent contradiction of simultaneous
presence of A-star and WR-star features in the spectrum. 

If the presence of an A-type star is not confirmed, however, it may appear
that SS~433 really is a WR X-ray binary, the second known in the Galaxy after Cyg~X-3.

\section{Conclusions}
\label{sec:Conclusions}
We find in Sect.~\ref{sec:population} that in principle there are two possible
He-star binary configurations which may explain the observed $\dot{P}/P$ and
$\dot{M}$ of Cyg X-3, and for which population-synthesis calculations
 combined with disk-formation criteria
predict the existence of 
\rev{$\sim 1$}
such system in the Galaxy.

The first possibility is a system consisting of a massive ($\gtrsim 7 M_\odot$) helium (i.e., WR) star
and a BH around which a disk is formed through wind accretion.
Population synthesis predicts that at any time 
$\sim 1$ such system with an orbital period similar to that of Cyg~X-3 is
present in the Galaxy, provided that the wind
velocity near the orbit of the c.o. is $\lesssim 1000$ km s$^{-1}$. In this
case the system will have a lifetime of several times $10^5$ yrs (the lifetime
of the He star) and the secular orbital-period increase is simply due to
stellar-wind mass loss.

The second possibility is a system consisting of a He star with a mass $\lesssim 1.5 M_\odot$ and a NS,
which is powered by mass transfer due to RLOF, at a rate in the
range $(1 - 3) \times 10^{-6} M_\odot$ yr$^{-1}$.
Population synthesis predicts that 
also $\sim 1$ such system with $P < 10$ hrs 
may be present in the Galaxy at any time. In this case the system will have a
lifetime of order $10^5$ yrs and the secular orbital-period increase is due to
the combined effects of the mass transfer and subsequent mass loss
--- at
a rate close to the transfer rate --- from the accretion disk around the NS.

In view of the population-synthesis results, we deem both
configurations equally likely for Cyg~X-3. We note that, though the
first of the solutions implies the presence of a ``real'' WR star in
the system, its mass is probably not extremely high,  $\lesssim
12\,M_\odot$. Thus, both solutions are consistent with the conclusion
of a relatively moderate mass for the companion in Cyg~X-3, which
follows from the identification of the emission region with the vicinity of the compact object \citep{fender:1999,stark:2003}.
\\[0cm]

We now speculate on the fate of these configurations.

In the 
``wind'' case, if the WR star loses sufficient mass, it might terminate as a NS, such that a system consisting of a BH plus a young
radio pulsar would emerge. In view of the likely mass of $\sim 5 M_\odot$ for
the BH
\citep[a value well within the range of BH masses in confirmed BH binaries, see, e.g.,][]{mcclintock:2004}, disruption of the system in the supernova
explosion seems unlikely. Alternatively, the WR star might collapse to a black
hole, producing a double-BH binary.

The fate of the 
``RLOF'' configuration is most likely a binary consisting of a
massive white dwarf (composed of CO or ONeMg) together with a recycled neutron
star. Several such systems are known in our galaxy as fast-spinning binary radio
pulsars with massive white-dwarf companions in circular orbits \citep[see,
e.g.,][]{stairs:2004}.

\begin{acknowledgements}
We are grateful to P. Eggleton for providing the latest version of his evolutionary code.
We would like to thank our colleagues at the ``Anton Pannekoek''
Astronomical Institute and at the Institute of Astronomy of the
Russian Academy of Sciences for useful discussions.
In particular we would like to thank T. Maccarone for discussions on
X-ray binaries, J. Miller-Jones for sharing preliminary results on
Cyg~X-3, and K. van der Hucht, A. 
de Koter, and N.N. Chugai for
useful discussions on the properties of WR stars.
LRY acknowledges warm hospitality of the Astronomical  Institute ``Anton
Pannekoek'' where part of this work was accomplished. LRY is supported by RFBR
grant 03-02-16254, the Russian Ministry of Science and Education Program ``Astronomy'', NWO, and NOVA. SPZ is supported by KNAW.

\end{acknowledgements}

\begin{center}
  APPENDIX A
\end{center}

We derive below the dependence of $\dot{P}/P$ on the masses of the
donor and the accretor $m_{\rm d}$ and $m_{\rm a}$, the mass-loss rate from the donor
$\dot{m}_{\rm d}$, the fraction of $\dot{m}_{\rm d}$ that goes in a direct wind $\alpha$,
and 
the fraction of $\dot{m}_{\rm d}$ that is re-ejected after transfer to the accretor $\beta$.
Early similar derivations are given in e.g. \citet[][]{huang:1963} and
\citet[][]{tutukov:1971}.

The total orbital angular momentum of the system is
\begin{equation}\label{eq:total_orbital_angular_momentum}
	J = \frac{2 \pi}{P} \left[ \frac{m_{\rm a} m_{\rm d}}{(m_{\rm a}+m_{\rm d})} \right] a^2.
\end{equation}
From Eq.~(\ref{eq:total_orbital_angular_momentum}) we have
\begin{equation}
	\frac{\ud J}{J} = - \frac{\ud P}{P} + \frac{\ud m_{\rm a}}{m_{\rm a}} + \frac{\ud
m_{\rm d}}{m_{\rm d}} - \frac{\ud (m_{\rm a}+m_{\rm d})}{(m_{\rm a}+m_{\rm d})} + 2 \frac{\ud a}{a}.
\end{equation}
We then use Kepler's third law and 
definitions
\begin{equation}
	\ud m_{\rm a} = - (1-\alpha-\beta) \ud m_{\rm d} \label{eq:dm_a}
\end{equation}
\begin{equation}
	\ud (m_{\rm a}+m_{\rm d}) = (\alpha+\beta) \ud m_{\rm d} \label{eq:dm_a_plus_m_d}
\end{equation}
(by definition, $\alpha + \beta \le 1$) to derive
\begin{equation}\label{eq:derivative_of_P2}
	\frac{\ud P}{P} = 3 \frac{\ud J}{J} + 3 \frac{(1-\alpha-\beta) \ud
m_{\rm d}}{m_{\rm a}} - 3 \frac{\ud m_{\rm d}}{m_{\rm d}} + \frac{(\alpha+\beta) \ud m_{\rm d}}{(m_{\rm a}+m_{\rm d})}.
\end{equation}
The logarithmic derivative of the orbital angular momentum due to
mass loss from the system
is given by
\begin{equation}\label{eq:dJ_over_J}
	\frac{\ud J}{J} = \frac{\alpha m_{\rm a}^2 + \beta m_{\rm d}^2}{(m_{\rm a}+m_{\rm d})} \frac{\ud
m_{\rm d}}{m_{\rm a} m_{\rm d}}.
\end{equation}
Inserting  this
into (\ref{eq:derivative_of_P2})
we obtain
\begin{align}\label{eq:derivative_of_Pfinal}
  \frac{\ud P}{P} & = \frac{\ud m_{\rm d}}{m_{\rm a} m_{\rm d} (m_{\rm a} + m_{\rm d})} \times \nonumber\\
  & \times \left[ (3 m_{\rm a}^2 - 2 m_{\rm a} m_{\rm d} - 3 m_{\rm d}^2)\alpha - 2 m_{\rm a} m_{\rm d} \beta + 3 m_{\rm d}^2 - 3 m_{\rm a}^2 \right].
\end{align}

We note that our Eq.~\ref{eq:derivative_of_Pfinal} is not consistent
with the expressions for variations of $a$ and $P$ in a similar model given by 
\citet[][their Eqs.~30 and 31]{soberman:1997}. The correct equations in their notation are:
\begin{eqnarray}\nonumber
  \mbox{} \qquad \frac{\partial \ln{a}}{\partial \ln{q}} =  2(\mathscr{A}_w - 1)
+ (1 - 2 \mathscr{B}_w) \frac{q}{1 + q} \nonumber \\
  \mbox{} \qquad  + (3 + 2 \mathscr{C}_w) \epsilon \frac{q}{1 + \epsilon q},
\label{eq:soberman30} \\
  \mbox{} \qquad \frac{\partial \ln{P}}{\partial \ln{q}} =  3(\mathscr{A}_w - 1)
+ (1 - 3 \mathscr{B}_w) \frac{q}{1 + q} \nonumber \\
  \mbox{} \qquad  + (5 + 3 \mathscr{C}_w) \epsilon \frac{q}{1 + \epsilon q}.
\label{eq:soberman31}
\end{eqnarray}\nonumber

\begin{center}
  APPENDIX B
\end{center}

\begin{figure}[!ht]
	\resizebox{\hsize}{!}{\includegraphics[angle=0]{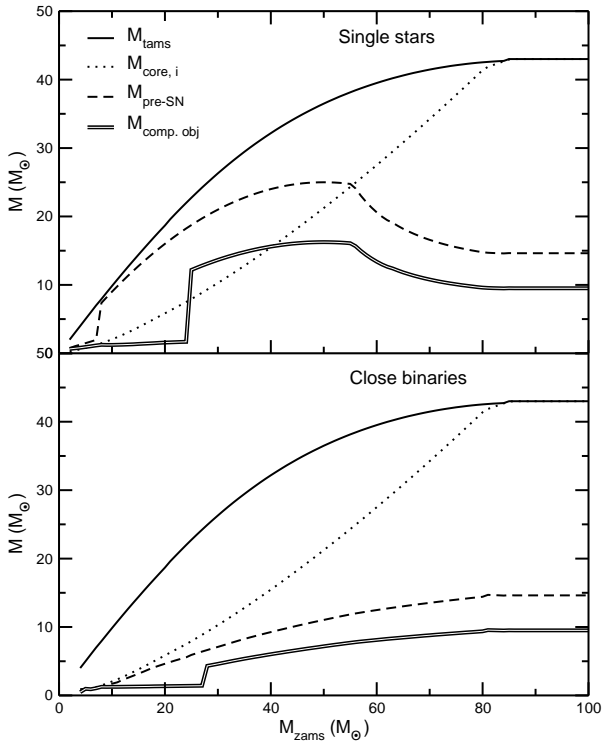}}
	\caption{Summary of the evolution of the mass of single stars  or components of wide binaries (top panel) and 
 of stars in binaries experiencing RLOF (bottom panel). As a function of initial mass the lines give 
respectively: the mass at the end of the main-sequence ($M_{\rm tams}$, solid 
line), the initial mass of the He-core ($M_{\rm core, i}$, dotted line), 
the mass just before the supernova explosion or the formation of the white 
dwarf ($M_{\rm pre-SN}$, dashed line) and the mass of the final c.o. 
($M_{\rm comp. obj}$, black-white-black line). 
The bottom panel
shows the masses for a primary that loses its hydrogen envelope soon after the 
end of the main sequence, before the He-core burning starts.}
	\label{fig:XXX}
\end{figure}

We used the \citet{nelemans:2004} realisation of the \textsf{SeBa} program \citep[][]{portegieszwart:1996} to carry out our population-synthesis calculations.
The effects of stellar 
wind and 
c.o. formation are shown in Fig.~\ref{fig:XXX}. The most 
important assumptions about the evolution of  massive stars in binaries 
relevant for this paper are as detailed in \citet{portegieszwart:1998,
  portegieszwart:1999, nelemans:2001b} with the following two differences:\\
 (1) WR stellar winds 
are modelled by the fit to observed WR mass-loss rates 
derived by \citet[][]{nelemans:2001}.\\
\noindent
(2) It is assumed that stars with carbon-oxygen 
cores more massive than 6.5 $M_\odot$ collapse into a BH (for single stars 
this means stars with an initial mass above 25 $M_\odot$, in close binaries stars 
with an initial mass above 28 $M_\odot$, see Fig.~\ref{fig:XXX}). The mass of the resulting 
BH is 65 per cent of the mass of the exploding object \citep[based on, e.g.,][]{nelemans:1999}. The lowest-mass BH (for an 
exploding naked carbon-oxygen core)  thus has a mass of 4.2 
$M_\odot$.

The bottom panel of Fig.~\ref{fig:XXX} summarises the evolution
  of binaries in which RLOF occurs. For which period this happens is a
  function of the mass of the primary.
As can be seen
from the figure we skip RLOF for any case-B period for initial masses $> 80 M_\odot$. Note that RLOF may be omitted for
much lower primary masses due to the occurrence of an LBV phase \citep[e.g.,][]{vanbeveren:1998b}. However, a detailed discussion
of this matter is beyond the scope of this paper.

\pagebreak

\bibliographystyle{aa}
\bibliography{references}

\end{document}